\renewcommand{\a}{\alpha} \renewcommand{\b}{\beta}
\newcommand{\g}{\gamma}
\renewcommand{\d}{\delta}  
  
\renewcommand{\l}{\lambda} \newcommand{\m}{\mu} \newcommand{\n}{\nu}
 \newcommand{\p}{\pi} 
 \renewcommand{\t}{\tau} 
  
\newcommand{\G}{\Gamma}  
  \renewcommand{\P}{\Pi}
 \newcommand{\Ph}{\Phi} 

\newcommand{\pr}{\prime}

\input epsf
\documentstyle[12pt]{article}

\begin{document}

\begin{titlepage}
\renewcommand{\baselinestretch}{1.3}
\hfill TUW - 93 -07\\
\medskip

\begin{center}
{\Huge Analytic solutions for cosmological perturbations in
multi-dimensional space-time
\footnote{Talk given at the Journ\'ees Relativistes '93, 5 -- 7
April, Brussels, Belgium}
\footnote{Work supported by the Austrian ``Fonds zur F\"orderung
der wissenschaftlichen Forschung'' under contract no.~P9005-PHY.}}
\bigskip

{\large Dominik J. SCHWARZ
\footnote{\tt email-address: dschwarz@email.tuwien.ac.at}}
\medskip

Institut f\"ur Theoretische Physik \\
Technische Universit\"at Wien \\
Wiedner Hauptstra\ss e 8-10, A-1040 Wien \\
Austria
\end{center}

\bigskip

\begin{abstract}
We obtain analytic solutions for the density contrast and the
anisotropic pressure in a multi-dimensional FRW cosmology with
collisionless, massless matter. These are compared with
perturbations of a perfect fluid universe.  To describe the
metric perturbations we use manifest gauge invariant metric
potentials. The matter perturbations are calculated by means of
(automatically gauge invariant) finite temperature field theory,
instead of kinetic theory.
\end{abstract}
\bigskip

May, 1993 \hfill
\end{titlepage}

The study of multi-dimensional cosmological models \cite{mdc}
may give some hints to explain the existence of just four
visible space-time dimensions. Contracting internal dimensions
(Kaluza-Klein cosmologies) could produce observable effects in
our visible space-time, e.g.\ gravity waves \cite{gas} or a de
Sitter phase (producing ``enough'' inflation in $D\,
{\raisebox{+ 2.5pt}{$>$} \hspace{-9pt}
\raisebox{-2.5pt}{$\sim$}}\, 40$ \cite{abb}).

Cosmological perturbations \cite{lif} are believed to provide
the seeds for the large scale structure in our universe. The
task of this work is to find out how the evolution of
cosmological perturbations depends on the dimension of the
universe.

We consider a radiation dominated, spatially flat
Friedman-Robertson-Walker (FRW) background in arbitrary
dimension $D \geq 4$. This background is chosen for reasons of
simplicity. Cosmological perturbations for collisionless,
massless matter are calculated and compared with perfect fluid
perturbations.  In contrast to this work, Fabris and Martin
\cite{fab} dealt with perturbations of Kaluza-Klein models
without matter in their contribution to this year's Journ\'ees
Relativistes.  Our units are fixed by $\hbar = c = k_B = 1$.

Cosmological perturbations can be classified into scalar, vector
and tensor perturbations \footnote{This classification is due to
space-coordinate transformations on the constant time
hypersurface.}.  An appropriate tool for describing the
corresponding metric perturbations $\d g_{\m\n}(x)$ is provided
by Bardeen's \cite{bar} gauge invariant metric potentials
\footnote{For a recent review see Ref.~\cite{muk}.}. In the
following, we only deal with scalar perturbations. Their two
metric potentials will be denoted by $\Ph$ and $\P$ (their
definition differs from that of Bardeen's $\phi_A$ and
$\phi_H$). They are related to the density contrast $\d$ and the
anisotropic pressure $\p_{anis}$, i.e.
\begin{eqnarray}
\label{d}
\d(x) = \frac{(D-2)^2}{4}\frac{x^2}{D-1}\Ph(x) \\
\label{p}
\p_{anis}(x) = \frac{D-2}{2}\frac{x^2}{D-1} \P(x) \, ,
\end{eqnarray}
on a space-like hypersurface representing the local restframe of
matter  everywhere.  Here $x=k\t$, where $k$ denotes the
wavenumber of the perturbations and $\t$ the conformal time of
the FRW background.

The evolution of $\Ph$ and $\P$ is determined by the
Einstein-Jacobi equations
\[
\d\left(\sqrt{-g}G_{\m\n}\right) =
8\p G\, \d\left(\sqrt{-g}T_{\m\n}\right)\, ,
\]
or equivalently
\begin{equation}
\label{ej}
\d\left(\sqrt{-g}G_{\m\n}\right) = 16\p G\, \d\left(
\frac{\d \G^M}{\d g^{\m\n}}\right)\, .
\end{equation}
$G_{\m\n}$ is the Einstein tensor and $T_{\m\n}$ the energy
momentum tensor. $\G^M$ denotes the matter part of the effective
action.

The calculation of the r.h.s.\ of Eq.~(\ref{ej}) is done by
means of finite temperature field theory. This method was
developed by Kraemmer and Rebhan \cite{kra}. The second
variation of the matter effective action (the r.h.s.\ of
Eq.~(\ref{ej})) is the contribution of gravitationally coupled
matter to the graviton self-energy. Here we restrict ourselves
to the case of {\em collisionless}, massless matter, e.g.\
background gravitons and neutrinos after their decoupling.
To calculate perturbatively the graviton self-energy
\begin{equation}
\label{pi}
\frac{\d^2 \G^M}{\d g^{\a\b}(x) \d g^{\m\n}(y)} =:
\sqrt{-g}\P_{\a\b\m\n}(x,y)
\end{equation}
we keep the temperature $T$ below the Planck scale, i.e.\
$T^{D-2}\ll G^{-1}$. From the Einstein equation and the
radiation energy-momentum tensor $T_{\m\n} \sim T^D$ we estimate
$R_{\m\n} \sim GT^D \ll T^2$. Thus curvature terms are of
subleading order. On the other hand, we assume that the
temperature is much higher than the incoming momenta (high
temperature expansion). So only high temperature 1-loop
contributions ($ \sim T^D$) to (\ref{pi}) have to be calculated.
This has been done in Ref.~\cite{reb} for $D=4$. The advantage
of the field theory approach --- instead of solving the
Boltzmann equation --- is the guaranteed gauge invariance.

Having calculated $\P_{\a\b\m\n}$ we have to solve the
Einstein-Jacobi equations (\ref{ej}) for $\Ph$ and $\P$. They
read:
\begin{eqnarray}
\label{tr}
\lefteqn{
\Ph^{\pr\pr}(x) + {4\over x}\Ph^{\pr}(x) + {1\over D-1}\Ph(x) =
{2\over D-1}\P(x) - {4\over D-2}{1\over x}\P^{\pr}(x)
} \\
\lefteqn{
\left({D-2\over 2}x^2 + D - 1\right)\Ph(x) +
\left(D-1\right)x\Ph^{\pr}(x) =
} \nonumber \\
& & 2(D-1)\left({D-2\over 2}\Ph +\P\right)(x) -
    {4(D-1)D\over \sqrt{\p}(D-2)}
    \G\left(D-1\over 2\right) \times
\nonumber \\
& & \left[
\int^x_0 \left(2\over x-x^{\pr}\right)^{D-4\over 2}
\sqrt{\p\over 2(x-x^{\pr})}\, {\rm J}_{D-3\over 2}(x-x^{\pr})
\left({D-2\over 2}\Ph +\P\right)^{\pr}\!(x^{\pr})\  dx^{\pr} +
\right.
\nonumber\\
\label{00}
& & \left. + \left(2\over x\right)^{D-4\over 2} \sqrt{\p\over 2x}
\sum_{n=0}^\infty \g_n {\rm J}_{D-3+2n\over 2}(x)
\right]
\end{eqnarray}
Eq.~(\ref{tr}) is the trace of Eq.~(\ref{ej}), whereas
Eq.~(\ref{00}) corresponds to its 0-0 component. The initial
conditions are implemented by fixing the arbitrary constants
$\g_n$ in Eq.~(\ref{00}). We simply set $\g_0 = const$ and $\g_n
= 0$ for $n\geq 1$. This choice corresponds to the initial
values:
$
{\P(0)\over \Ph(0)} = - {D-3\over D+3}\, .
$
In kinetic theory similar equations to (\ref{tr}) and
(\ref{00}) have been derived in particular gauges and studied in
Refs.~\cite{pee}. Solutions have been obtained numerically
only.

\begin{figure}[h]
\epsfbox{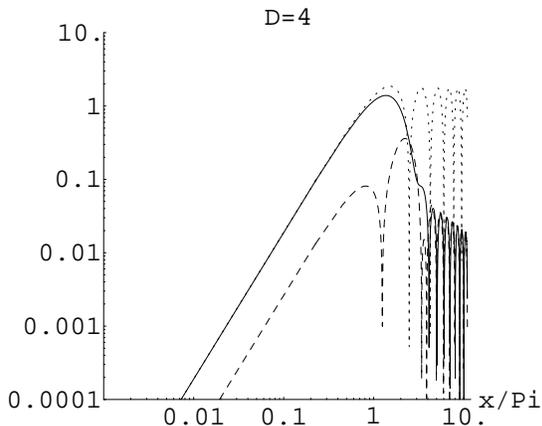}
\caption{The absolute values of the density contrast $|\d|$
(full line) and the anisotropic pressure $|\p_{anis}|$ are
plotted over $x/\p$ for a four dimensional FRW universe, matter
being massless and collisionless. The density contrast for
collisional matter $|\d_{pf}|$ is shown by the dotted line.}
\label{f1}
\end{figure}

\begin{figure}[h]
\epsfbox{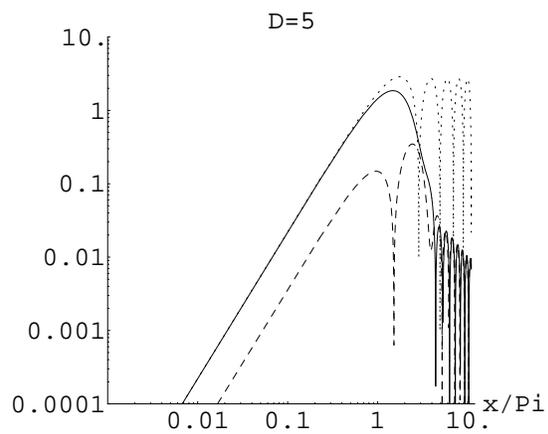}
\caption{As Fig.~1, but $D=5$.}
\label{f2}
\end{figure}

\begin{figure}[h]
\epsfbox{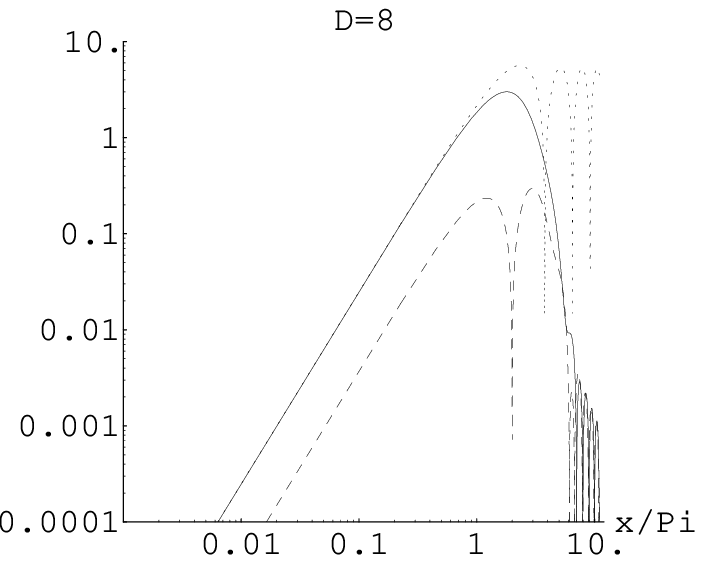}
\caption{As Fig.~1, but $D=8$.}
\label{f3}
\end{figure}

Exact analytic solutions to (\ref{tr}) and (\ref{00}) may be
found by a power series Ansatz
\cite{kra} yielding for the first terms:
\begin{eqnarray*}
\Ph(x) &=& const\left((D+3) -
{3D^3+11D^2-20D-30\over 5D^2-4D-10}{x^2\over 2!}
+ \cdots \right) \\
\P(x) &=& const\left(-(D-3) +
{(D-2)(11D^2-17D-30)\over 2(D+3)(5D^2-4D-10)}
{x^2\over 2!} -  \cdots \right)\, .
\end{eqnarray*}
The metric potentials can be calculated with arbitrary precision
since the series are rapidly converging for all $x$.  This
determines the density contrast and the anisotropic pressure
through Eqs.~(\ref{d}) and (\ref{p}). The results are plotted in
Fig.~\ref{f1} for $D=4$, in Fig.~\ref{f2} for $D=5$, and in
Fig.~\ref{f3} for $D=8$.

Density perturbations of {\em collisional},
massless matter, e.g.\ photons before recombination, are also
plotted in Figs.~\ref{f1} -- \ref{f3}. They are given by the
solution of Eq.~(\ref{tr}) with $\P$ vanishing.  This happens,
because the photons interact via Thomson scattering, and
therefore no anisotropic pressure can evolve. Matter remains a
perfect fluid.  The solution
$
\d_{pf}(x) \sim x{\rm j}_1({x\over\sqrt{D-1}})
$
has already been obtained in Ref.~\cite{kod}.

To discuss Figs.~\ref{f1} -- \ref{f3} note that ${x\over \p} =
({\l\over 2})^{-1} R_H$ ($R_H$ being the comoving horizon size).
Far outside the horizon (${x\over \p} \ll 1$) all perturbations
grow as $x^2$ --- independent of dimension. This growth is
related to the gravitational (Jeans) instability.

Well inside the horizon (${x\over \p} \gg 1$) the perturbations
oscillate. For perfect fluids their amplitude is constant. Their
wavelenght is determined by their dimension dependent sound
velocity ($v_S={1\over\sqrt{D-1}}$). For collisionless matter
the perturbations are damped ($\sim x^{- (D-2)/2}$) due to
directional dispersion.  This dimension dependent effect can be
understood easily \cite{boe}. In more than one dimension peaks
and throughs of waves travel in different directions, and
consequently the wave is damped. These waves are propagating
with the speed of light.

Another effect is observed in Figs.~\ref{f1} -- \ref{f3}. The
maximum of the perturbations is moving to higher values of $x$
with higher dimension.  The maximum value itself is fixed by the
arbitrary constant $\g_0$.

We showed that the method developed in Refs.~\cite{reb,kra,vec}
works in multi-di\-men\-sio\-nal space-time, too. Vector\footnote{A
detailled treatment of vector perturbations for $D=4$ is given in
Ref.~\cite{vec}.} and tensor perturbations could be calculated
in the same manner.  Two dimension dependent effects have been
demonstrated: Damping of collissionless perturbations is due
to directional dispersion, and perturbations start to ``feel''
the horizon at a later time in higher dimensional universes.

I would like to thank A.\ K.-Rebhan for a lot of discussions and
useful comments. Moreover, conversations with J.C.\ Fabris and
J.\ Martin have been very helpful.

\end{document}